\documentclass[letter]{aa} % for the letters 
\usepackage{hyperref}

\usepackage{xcolor}
\definecolor{dark-red}{rgb}{0.9,0.0,0.0}
\definecolor{dark-blue}{rgb}{0.15,0.15,0.9}
\definecolor{dark-green}{rgb}{0.15,0.8,0.15}
\definecolor{medium-blue}{rgb}{0,0,0.9}
\hypersetup{
colorlinks, linkcolor=red,
citecolor={dark-blue} , urlcolor={medium-blue}
}
\usepackage{changepage} % for \begin{adjustwidth}{-Xcm}{} \end{adjustwidth}

\usepackage{graphicx}
\usepackage{txfonts}
\begin{document} 

%6 M$_\mathrm{Jup}$ 

\title{Three planets around HD~27894} 
\subtitle{A close-in  pair with a 2:1 period ratio and 
an eccentric Jovian planet at 5.4~AU\thanks{Based on observations collected at 
the European Organisation for Astronomical Research in the Southern Hemisphere under ESO 
programmes 072.C-0488, 192.C-0852 and 097.C-0090.}}

%\vspace{-3mm}

\author{T. Trifonov\inst{1}, M. K\"{u}rster\inst{1}, M. Zechmeister\inst{2}, O. V. Zakhozhay\inst{1,3}, 
S. Reffert\inst{4}, M.~H. Lee\inst{5,6}, F. Rodler\inst{7}, S.~S. Vogt\inst{8}, \and S.~S.~Brems\inst{4}}

\institute{Max-Planck-Institut f\"{u}r Astronomie,\ K\"{o}nigstuhl  17, 69117 Heidelberg, Germany\\ 
\email{trifonov@mpia.de}
\and
Institut f\"{u}r Astrophysik, Georg-August-Universit\"{a}t, Friedrich-Hund-Platz 1, 37077, G\"{o}ttingen, Germany\\ \vspace{-3mm}
\and     
Main Astronomical Observatory, National Academy of Sciences of the Ukraine, Ukraine\\ \vspace{-3mm}
\and                 
Landessternwarte, Zentrum f\"{u}r Astronomie der Universit\"{a}t Heidelberg,\ K\"{o}nigstuhl 12, 69117 Heidelberg, Germany\\ \vspace{-3mm}
\and
Department of Earth Sciences, The University of Hong Kong, Pokfulam Road, Hong Kong   \\     \vspace{-3mm}      
\and        
Department of Physics, The University of Hong Kong, Pokfulman Road, Hong Kong\\ \vspace{-3mm}
\and         
European Southern Observatory (ESO), Alonso de Cordova 3107, Vitacura, Santiago de Chile, Chile\\ \vspace{-3mm}
\and         
UCO/Lick Observatory, University of California, Santa Cruz, CA, 95064, USA \\ \vspace{-3mm}
}

\date{Received 25 April 2017 / Accepted 30 May 2017}

% \abstract{}{}{}{}{} 
% 5 {} token are mandatory

\abstract
{}
{Our new program with HARPS aims to detect mean motion resonant planetary systems around 
stars which were previously reported to have a single bona fide planet, often based only on
sparse radial velocity data.
}
{Archival and new HARPS radial velocities for the K2V star HD~27894 
were combined and fitted with a three-planet self-consistent dynamical model.
%\LEt{ including three planets?  containing ? (to avoid repeating consistent/consisting) } 
The best-fit orbit was tested for long-term stability.}
{We find  clear evidence that HD~27894 is hosting at least three massive planets.
In addition to the already known Jovian planet with a period $P_{\rm b}$ $\approx$ 18 days we 
discover a Saturn-mass planet with $P_{\rm c}$ $\approx$ 36 days, likely in a 2:1 mean motion resonance with the first planet, 
and a cold massive planet ($\approx$ 5.3 $M_{\mathrm{Jup}}$) 
with a period $P_{\rm d}$ $\approx$ 5170 days
on a moderately eccentric orbit ($e_{\rm d}$~=~0.39).
}
{HD~27894 is hosting a massive, eccentric giant planet orbiting around a 
tightly packed inner pair of massive planets likely involved in an asymmetric 2:1 mean motion resonance.
HD~27894 may be an important milestone for probing planetary formation and evolution scenarios.}


\authorrunning{T.~Trifonov et al.}
\titlerunning{Three planets around HD~27894} 
\maketitle

\section{Introduction}

The radial velocity (RV) technique is very successful in determining
the orbital architectures of multiple extrasolar planetary systems.
In some exceptional cases N-body modeling of precise Doppler data in 
resonant systems can even 
reveal the system's short-term dynamics and constrain the planetary true masses 
%\LEt{ one planet's mass; multiple planets' masses (NB position of ' ) } 
and mutual inclinations \citep[][]{Bean2009, Rivera2010,Trifonov2014}.
Therefore, RV multiple planetary system discoveries are fundamentally important 
in order to understand planet formation and evolution in general.
% However, many planets in multiple systems have been announced based on sparse RV data
% and sometimes incomplete phase coverage, leading to ill-determined system architectures. 
Many RV planet discoveries, however, have been announced based on sparse data samples,
and sometimes incomplete phase coverage leading to ill-determined system architectures. 
For example, the combined RV signal of a pair of planets in low eccentricity orbits 
near a 2:1 mean motion resonance (MMR) can be misinterpreted as a single planet 
with moderate eccentricity if the data are sparse \citep{Escude2010,Wittenmyer2013,Kurster2015}. 
In addition, the detection of long-period massive planets needs 
continuous precise RV measurements taken over a sufficient temporal baseline.

An example is the moderately hot ($a$ = 0.125 AU, $P \approx$ 18 days) Jovian planet 
HD~27894 b \citep{Moutou2005}, which was discovered based on only 
twenty RV measurements taken with the ESO HARPS spectrograph 
\citep[La Silla Observatory, Chile,][]{Mayor2003}.
Recently, \citet{Kurster2015} illustrated how the sparse 
RV data for HD~27894 can actually be modeled well 
with an additional Neptune-mass planet in an inner orbit near the 2:1 resonance with the known Jovian planet.
Later we found that an outer Saturn-mass planet
with a period of $\approx$~36~days provides an even better fit to the data.

We included HD~27894 in our HARPS RV monitoring program following up 
35~stars that were previously reported to harbor a single planet in order to look for additional planets.
The additional HARPS observations of HD~27894 confirm the existence of planet~c with a 36-day period, 
and reveal an additional long-period RV signal consistent with a 
massive Jovian planet ($m~\sin i$ = 5.4 $M_{\mathrm{Jup}}$, planet~d)
with a period of $\approx$ 5200~days. 
In this paper we present an updated orbital configuration for the HD~27894
system; according to our self-consistent N-body analysis of the RV data,
the system is composed of three planets, the inner two likely in a 2:1 MMR.

The paper is organized as follows. In Sect.~\ref{sec2} we 
start with an overview of the HARPS measurements and in Sect. \ref{sec2.1} we present 
results from our dynamical modeling, which reveals the three-planet system.  
In Sect.~\ref{sec3} we present a long-term dynamical analysis
of HD~27894 and we discuss the possible 2:1 MMR between planets b and c.
In Sect. \ref{sec4} we discuss the unusual orbital architecture of the HD~27894 system.

%-----------------------------------------------------------------

\begin{figure}[btp]
\begin{center}$
\begin{array}{cc} 

\includegraphics[width=8.6cm]{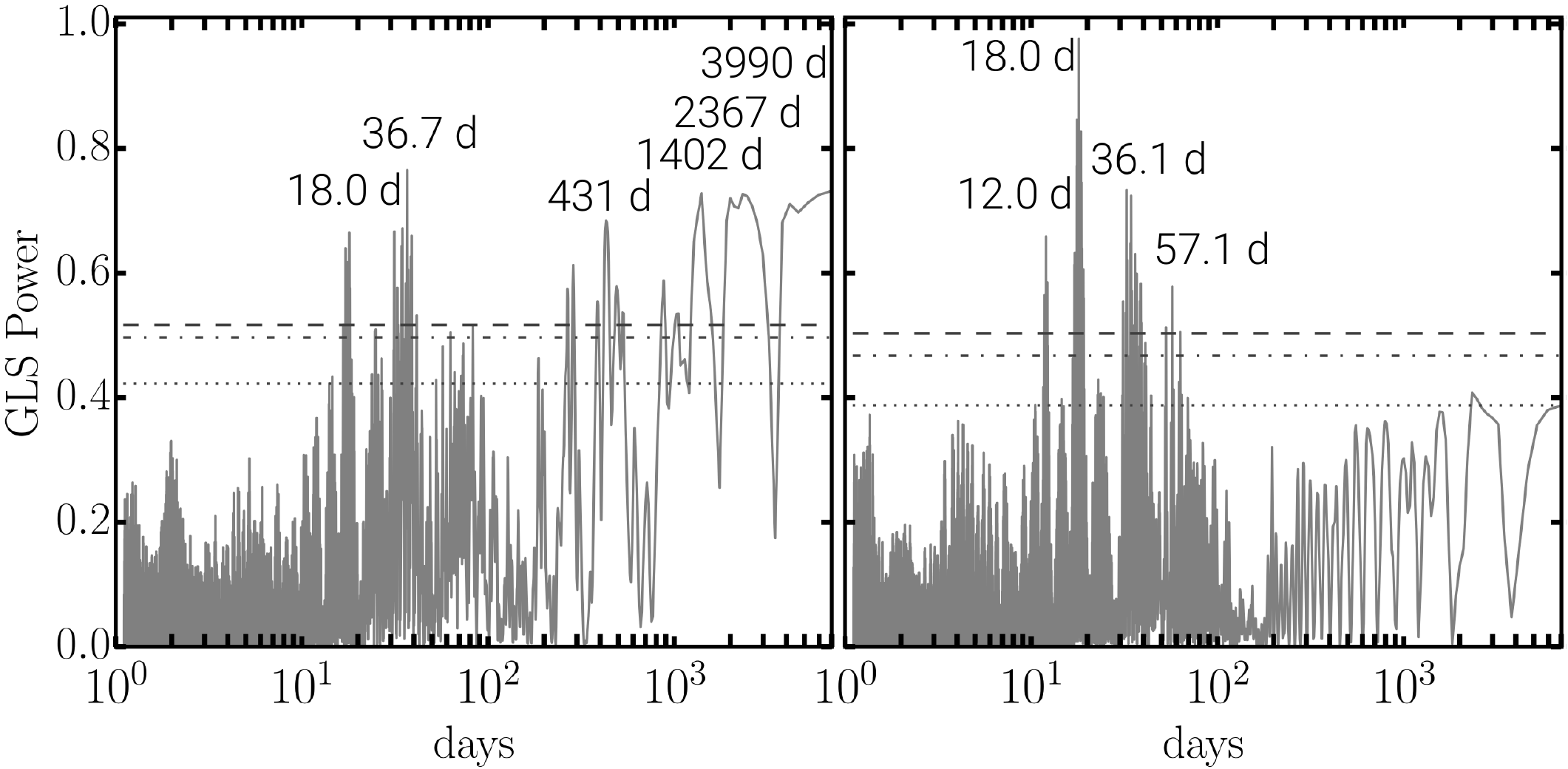} \put(-225,100){a)}  \put(-110,100){b)} \\
\includegraphics[width=9cm]{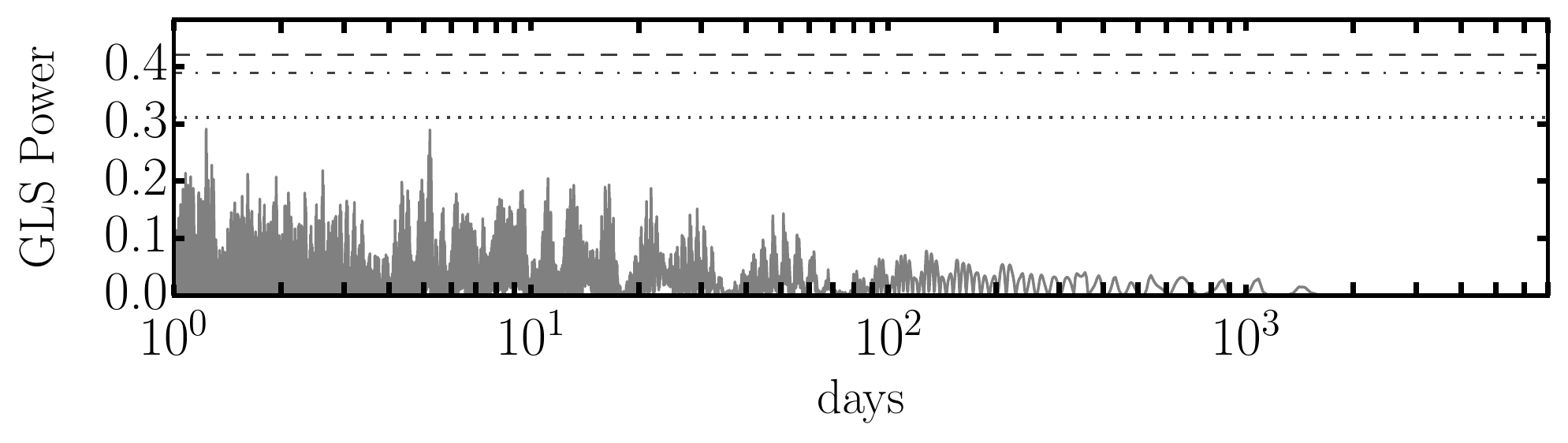}
\put(-215,57){c)}  \\

% % % \put(-15,57){\scriptf FAP < 1\%}

\end{array} $

\end{center}

\caption{GLS periodograms of the RV data of HD~27894 with FAP levels of 10\%, 1\%, and 0.1\%
calculated from a bootstrap analysis.
Panel a): The HARPS RVs resulted in many significant peaks especially
at long periods. Panel b): After removal of the RV signal from planet d ($P_{\rm b}$ = 5174 days),
the dominant peaks left are at 18.0 and 36.7~days (due to planets~b and c), and at 12.0 and 57.1~days (aliases).
Panel c): No significant peaks are left after fitting a 3PDM (see Table~\ref{table:multi}).
}
\label{gls}
\end{figure}

\section{HARPS data for HD~27894}
\label{sec2}

HD~27894 is a non-active solar-type star of spectral type K2V with an estimated 
mass of 0.8~$M_{\odot}$ and metallicity of $[Fe/H] = 0.30~\pm~0.07$ dex \citep{Moutou2005}.
The star is an ideal HARPS target, due to 
its low declination ($\delta$ = $-$59$^\circ$ 24' 41.40''), visual brightness of V = 9.36 mag, and
slow rotation period of $P_{\rm rot}$ $\sim$ 44~days \citep{Moutou2005}.

\begin{figure}[btp]
\begin{center}$
\begin{array}{cc} 

\includegraphics[width=4.5cm,height=3.7cm]{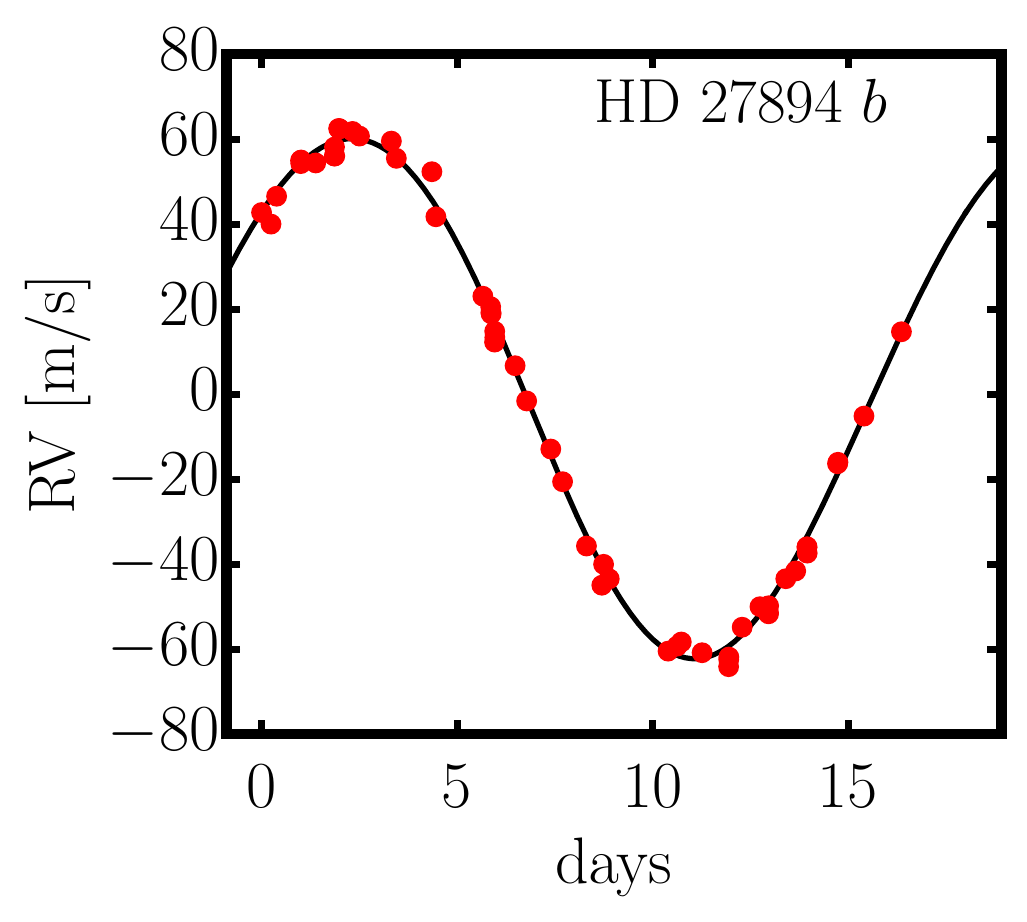} 
\put(-95,35){a)}  
\includegraphics[width=4.5cm,height=3.7cm]{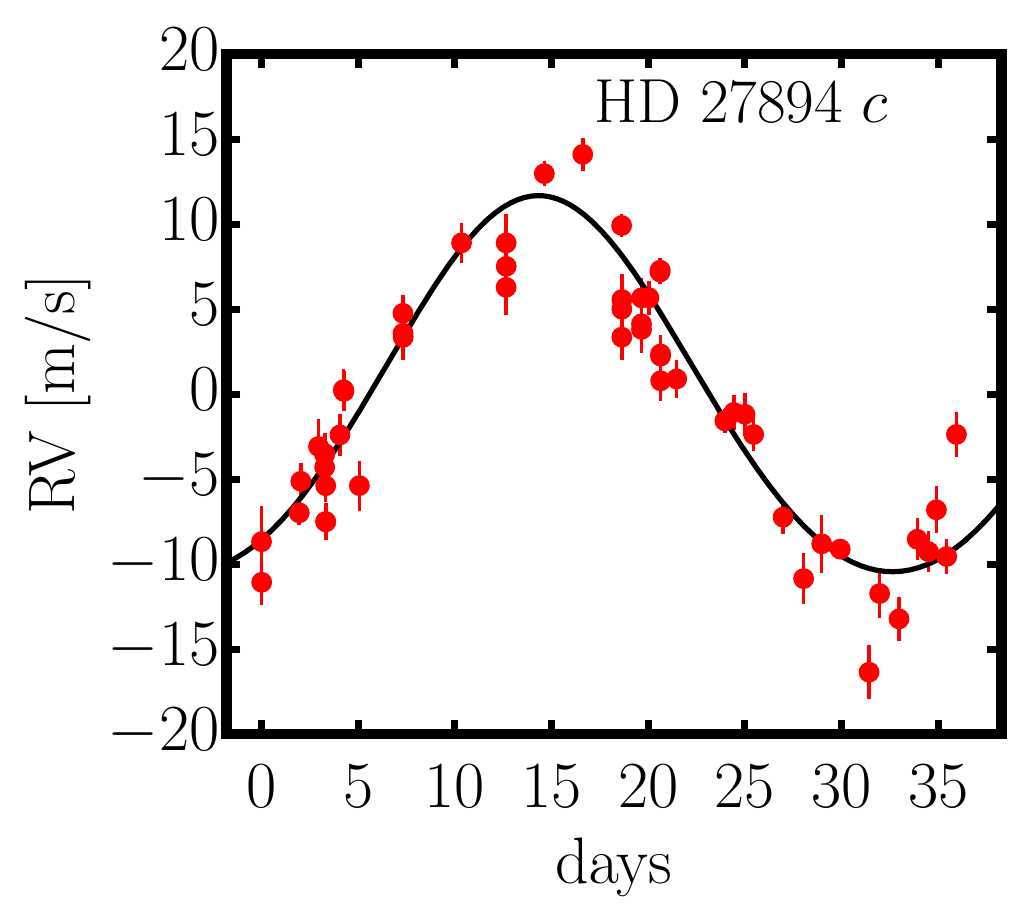}
\put(-95,85){b)}\\
\includegraphics[width=4.5cm,height=3.7cm]{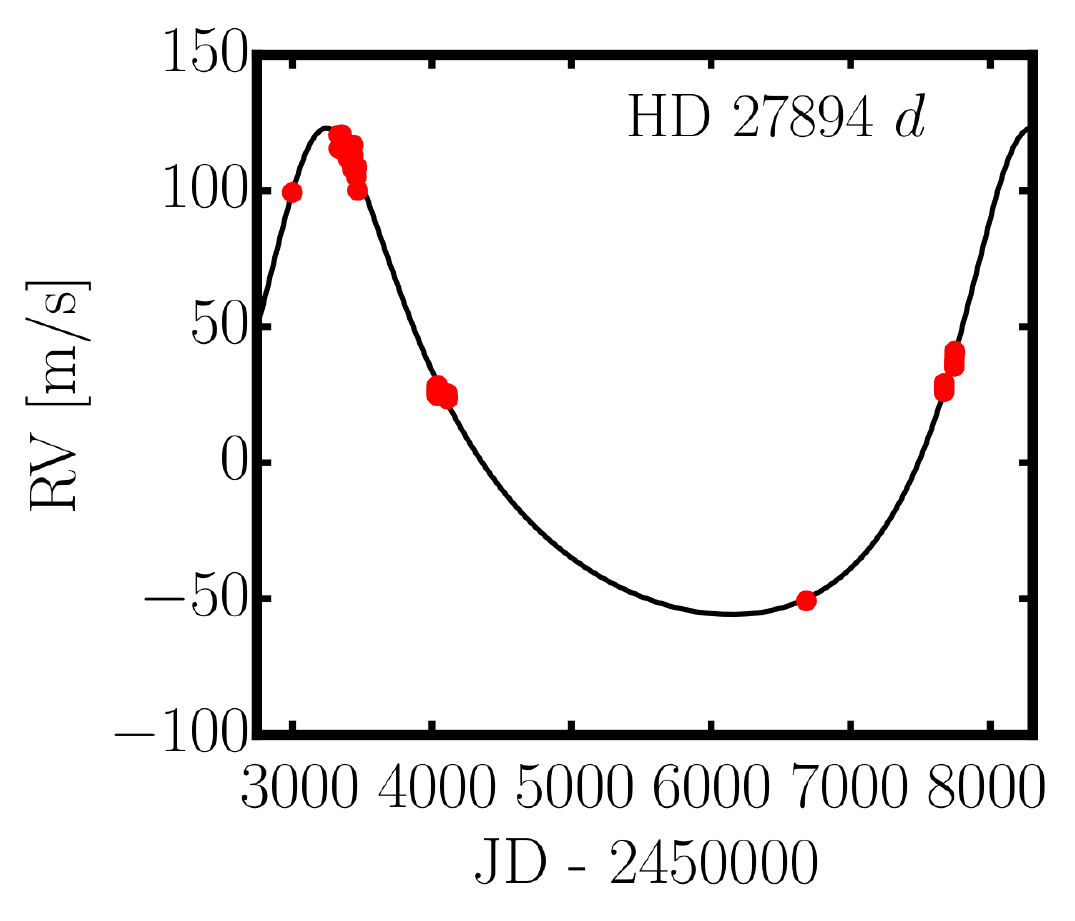}
\put(-95,35){c)} 
\includegraphics[width=4.5cm,height=3.7cm]{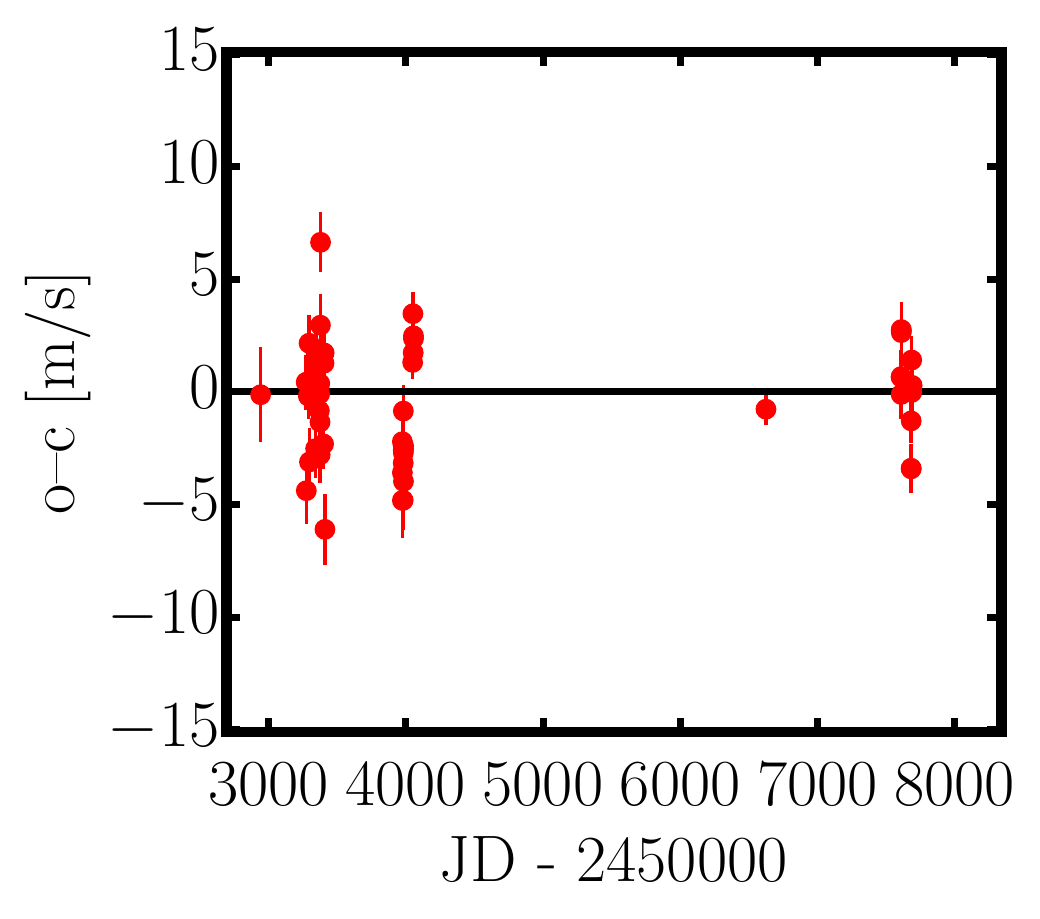}  
\put(-95,85){d)}\\

\end{array}$

\end{center}

\caption{Panels a) and b) illustrate the 3PKM signals of planets b and c,
phase folded to their periods of about 18 and 36 days, respectively. 
Panel c) illustrates the Doppler signal induced by planet~d,
and panel d) shows the residuals of the 3PKM.
}
\label{dynfit}
\end{figure}     

Based on the first twenty HARPS RV measurements taken from 2003 to 2005
\citet{Moutou2005} announced HD~27984~b, 
a Jovian planet with a minimum mass of 0.62~$M_{\mathrm{Jup}}$, orbiting at a semi-major axis of
0.125 AU in a nearly circular orbit (eccentricity $0.049 \pm 0.008$).
Subsequently, an additional 21 HARPS RVs were obtained in 2006, and one more in 
2013. Thus, we found a total of 42 HARPS spectra in the ESO archive. 
The archival RVs already clearly showed that HD~27984~b is not alone, and so  we 
included the star in our ongoing HARPS program aimed at finding hidden multiple planetary systems, 
and obtained 16 more RV measurements in 2016.

%The RVs for all measurements were recomputed with the SERVAL pipeline 
All RVs were recomputed with the SERVAL pipeline
(Zechmeister et al., in prep.), which was originally developed for the CARMENES project \citep{quirrenbach16}.
It includes a ``shifting and co-adding'' $\chi^2$ fitting algorithm,
where one of the fitting parameters is the RV \citep[][]{Escude2012}. 

The latest HARPS measurements were taken after the upgrade of the HARPS fibers in May 2015, which
introduces a notable RV offset between data taken before and after the upgrade.
\citet[][]{LoCurto2015} investigated the typical offsets introduced for various spectral
types, ranging from $-2$~m\,s$^{-1}$ for M~stars up to 20~m\,s$^{-1}$ for F~stars. 
For K~stars such as HD~27894, the mean offset value is 13.4~m\,s$^{-1}$, with a rather
small dispersion of only 2.5~m\,s$^{-1}$ among the five K~stars. 
We thus subtracted this offset value from the post-2015
HARPS radial velocities in order to put them on the same scale as the older velocities. 
We also tried to fit for this RV offset, 
but it turns out that it is not well constrained by our HARPS data set of HD 27894 alone.

\begin{table}[tp]

\centering  

%\resizebox{0.65\textheight}{!}{\begin{minipage}{\textwidth}

\caption{Best edge-on coplanar 3PDM for HD 27894.}   
\label{table:multi}

% \begin{adjustwidth}{0.5cm}{} 

\resizebox{0.33\textheight}{!}{\begin{minipage}{\textwidth/2} 

\begin{tabular}{l r r r r r r r r r}     
\hline \hline \noalign{\vskip 0.7mm}  

Orb. Param.    & planet b & & planet c & & planet d     \\     
\cline{1-6}
\noalign{\vskip 0.9mm}

$K$  [m\,s$^{-1}$]                 & 59.80$_{-0.59}^{+0.80}$   & & 11.57$_{-2.89}^{+0.79}$     & & 79.76$_{-15.03}^{+2.62}$     \\ \noalign{\vskip 0.9mm}
$P$ [days]                         & 18.02$_{-0.02}^{+0.01}$   & & 36.07$_{-0.09}^{+0.26}$     & & 5174$_{-82}^{+171}$          \\ \noalign{\vskip 0.9mm}
$e$                                & 0.047$_{-0.008}^{+0.012}$ & & 0.015$_{-0.002}^{+0.020}$   & & 0.389$_{-0.030}^{+0.087}$    \\ \noalign{\vskip 0.9mm}
$\varpi$ [deg]                     & 132.2$_{-15.2}^{+2.8}$    & & 44.2$_{-7.5}^{+11.0}$        & & 353.9$_{-3.1}^{+12.3}$      \\ \noalign{\vskip 0.9mm} 
$M$ [deg]                          & 193.9$_{-7.8}^{+12.6}$    & & 117.7$_{-4.2}^{+19.1}$       & & 343.8$_{-7.0}^{+2.5}$       \\ \noalign{\vskip 0.9mm}

$a$ [AU]                           & 0.125$_{-0.0001}^{+0.0001}$  & & 0.198$_{-0.001}^{+0.001}$ & & 5.448$_{-0.058}^{+0.119}$   \\ \noalign{\vskip 0.9mm}
$m$        [$M_{\mathrm{Jup}}$]    & 0.665$_{-0.007}^{+0.009}$    & & 0.162$_{-0.040}^{+0.011}$ & & 5.415$_{-1.214}^{+0.239}$   \\ \noalign{\vskip 0.9mm}
\cline{1-6} 
\noalign{\vskip 0.9mm}
\makebox[0.1\textwidth][l]{RV$_{\mathrm{off}}$~=~-86.9$_{-1.7}^{+13.2}$ m\,s$^{-1}$,~~~~$rms$~=~2.04 m\,s$^{-1}$,~~~~$\chi_{\nu}^2$~=~3.62} \\

\noalign{\vskip 0.5mm}
\cline{1-6} 

\end{tabular} 
\end{minipage}}

% \end{adjustwidth}
\end{table}

\section{RV analysis}
\label{sec2.1}

In Fig.~\ref{gls} we show the generalized Lomb-Scargle peridogram \citep[GLS;][]{Zechmeister2009} 
of the available HARPS data with false alarm probability (FAP) levels of 10\%, 1\%, and 0.1\%
calculated by randomly reordering 1000 bootstrap copies of the RV data \citep{Kuerster1997}.
We find many significant peaks, the most significant ones around 18 and 36~days (due to planets b and c) 
and several longer periods at $\sim$ 431, 1402, 2367, and 3990 days, one of them possibly due to planet d; 
the rest are likely aliases.
When it is assumed that the HD~27894 system is composed only of  planets b and d, 
our best two-planet Keplerian model (2PKM) yields reduced $\chi^2$ ($\chi_{\nu}^2$) = 8.89 and $rms$ = 3.05 m\,s$^{-1}$;
periods $P_{\rm b,d}$ = 18.02, 5067 days; eccentricities $e_{\rm b,d}$ = 0.04, 0.34; 
and minimum masses $m_{\rm b,d} \sin i$ =  0.64 and 6.20 $M_{\mathrm{Jup}}$, respectively, 
and represents a very strong minimum in the least-constrained $P_{\rm d}-e_{\rm d}$ parameter space.
Indeed, an outer massive companion with a $\sim$5000 day orbit and moderate eccentricity explains the data well since 
subtracting its RV signal from the data also removes the power at smaller periods, 
except the peaks near 18 and 36 days and a few others near 12 and 57 days.
The 12-day peak is a clear alias of two periods at 18 and 36~days (planets b and c),
while the 57-day peak is likely an alias of the
sampling frequency (likely related to the lunar cycle of $\sim$28 days)
and planet b. All significant peaks disappear when we remove planets b, c, and d (see~Fig.~\ref{gls}).

We thus fitted a three-planet Keplerian model (3PKM) to the RV data and obtained
periods $P_{\rm b,c,d}$ = 18.02, 36.44, 5051~days;
eccentricities of $e_{\rm b,c,d}$ = 0.03, 0.06, 0.39; 
and RV semi-amplitudes of $K_{\rm b,c,d}$ = 60.8, 11.1, 89.4 m\,s$^{-1}$, 
consistent with $m_{\rm b,c,d} \sin i$ =  0.67, 0.16, and 6.02 $M_{\mathrm{Jup}}$ 
and orbital semi-major axes of $a_{\rm b,c,d}$ = 0.125, 0.199, and 5.363~AU.
The $\chi_{\nu}^2$ of the fit corresponds to 6.18, and the $rms$ value to 2.55 m\,s$^{-1}$, 
which represents a significant improvement over the 2PKM.
The 3PKM Doppler signals are illustrated in Fig.~\ref{dynfit}.

As a next step we adopted the 3PKM best-fit parameters as an initial guess for a self-consistent three-planet dynamical model (3PDM),
which optimizes the planetary parameters by integrating the equations of motion
in Jacobi orbital coordinates \citep[e.g.,][]{LeeM2003, Tan2013}.
This model provides more realistic orbital parameter estimates than the unperturbed Keplerian model because it 
takes into account the gravitational interactions between the bodies while fitting the Doppler data. 
As we will show, this is necessary for our proposed  three-planet system around HD~27894,
given the large minimum masses and small period differences of planets b and c.
For our dynamical fitting we assumed only coplanar and edge-on fits
($i$ = 90$^\circ$); the investigation of other orbital architectures is beyond the scope of the current paper.

The 3PDM resulted in an even smaller $\chi_{\nu}^2$ of 3.62,
\footnote{$\chi_{\nu}^2$ above unity indicates the presence of
 additional data noise, e.g., due to stellar RV ``jitter'', mutual inclinations, or calibration errors.
For the 3PDM this noise is on the order of $\sim$ 1.5 m\,s$^{-1}$,
but its inclusion in the RV error budget does not have a significant impact on our analysis.}
and $rms$ value of 2.04 m\,s$^{-1}$.
For the outermost companion we derive a period of $P_{\rm d}$ = 5174 d ($a_{\rm d}$ = 5.45 AU), dynamical 
mass of $m_{\rm d}$ =  5.4 $M_{\mathrm{Jup}}$, and moderate eccentricity of $e_{\rm d}$ = 0.389. 
The inner planets have  $P_{\rm b}$ = 18.02 days and $P_{\rm c}$ = 36.07 days, low eccentricities 
of $e_{\rm b}$ = 0.05 and $e_{\rm c}$ = 0.015, and 
minimum dynamical masses of $m_{\rm b}$ = 0.664 $M_{\mathrm{Jup}}$ and $m_{\rm c}$ = 0.162 $M_{\mathrm{Jup}}$,
suggesting a Jupiter- and a Saturn-mass planet interacting gravitationally, possibly in a 2:1 MMR.
The 3PDM best-fit parameters and their uncertainties valid for JD = 2452941.826 are listed in Table~\ref{table:multi}. 
The asymmetric uncertainties of the orbital parameters were estimated by drawing 5000 
model independent bootstrap synthetic samples \citep[e.g.,][]{Press}, fitted with a 3PDM.

\subsection{Significance of the three-planet dynamical model}

The RV signals from planets b and d are large and can be well explained with periodic Keplerian motion.
Planet c, however, has $P_{\rm c} \approx 2P_{\rm b}$ and  $m_\mathrm{c} \approx {1\over 4} m_\mathrm{b}$,
thus yielding a smaller RV semi-amplitude (11.57~m\,s$^{-1}$) than its more massive neighbors.
Since we still have relatively sparse RV data to strongly constrain a
three-planet system, we need to exclude the possibility that the supposed planet~c is not real.

In order to convince ourselves of the existence of planet~c, we qualitatively compared a 
two-planet dynamical model (2PDM) consisting only of planets b and d with our 3PDM. 
While the best 2PDM (without planet c) yields $\chi_{\nu}^2$ = 8.89 and $rms$ = 3.05 m\,s$^{-1}$, 
a 3PDM (including planet c) yields $\chi_{\nu}^2$ = 3.62 and $rms$ = 2.04 m\,s$^{-1}$, 
which -- according to the F-Test -- is a significantly better fit (FAP = 9.1x$10^{-8}$).
Similarly, the best 3PKM with the same number of fitting 
parameters as the 3PDM has  $\chi_{\nu}^2$ = 6.18 and $rms$ = 2.55 m\,s$^{-1}$,
showing that including the dynamical interactions between planets b and c indeed leads to a better fit.

\vspace*{-0.1cm}

\section{System long-term dynamics}
\label{sec3}

The long-term dynamical analysis of our three-planet model was carried out
using the {\it SyMBA} symplectic integrator \citep{Duncan1998},
modified to work directly with the best-fit Jacobi orbits.
We integrated the system for 10~Myr with a step of
0.1 days to ensure accurate simulation and high orbital phase resolution.
We monitored the evolution of the planetary 
semi-major axes $a_{\rm b}$, $a_{\rm c}$, $a_{\rm d}$ and eccentricities $e_{\rm b}$, $e_{\rm c}$, $e_{\rm d}$ 
as a function of time to assure that the system remains
regular and well separated at any given time of the simulation.
Any deviation of the planetary semi-major axes by more than 20\% from
their starting values, or eccentricities leading to crossing orbits, were considered unstable.

\subsection{Stability of the three-planet system}

% We find that the HD~27894 system with its three planets is generally stable for 10~Myr.
We find that the HD~27894 system is generally stable for 10~Myr.
The orbital evolution is regular with no significant configuration anomalies.
The massive outer planet HD~27984~d experiences almost no change in its 
orbital separation and eccentricity, staying at  $a_{\rm d}$ =  5.4~AU, $e_{\rm d}$ = 0.39
through the course of the integration. 
Owing to the large separation, the secular perturbation of the inner
pair of planets due to planet~d is  negligible. 

Planets b and c, however, are dynamically very active, which is expected 
given their large masses and small orbital separation.
They have a period ratio close to 2:1, and mean eccentricities of $e_{\rm b}$ = 0.036 and $e_{\rm c}$ = 0.047.
As a large-scale stability test we integrated our 5000 3PDM bootstrap fits for 10 k\,yr and and we find
overall similar orbital behavior. 
Thus, we conclude that the HD~27984 system is long-term stable.
%\LEt{ ... that HD27984 is a long-term stable system.=or= ... that HD27984 is characterized by long-term stability. ? (if "long-term stable" is standard, pls ignore my suggestions }

\subsection{A closer look at the inner pair}
\label{closer}

The period ratio close to 2:1 suggests that the inner planets b and c are in resonance.
A 2:1 MMR is the lowest order resonance characterized by the two resonance angles 
$\theta_1 =  \lambda_{\rm b} - 2\lambda_{\rm c} + \varpi_{\rm b},~~\theta_2 =  \lambda_{\rm b} - 2\lambda_{\rm c} + \varpi_{\rm c}$
\citep[where $\lambda_{\rm b,c}$ =  $M_{\rm b,c}$ + $\varpi_{\rm b,c}$ is the mean 
longitude of planet b and c, respectively; see][]{Lee2004}. At least one of these variables must librate 
to find the system in a 2:1 MMR.
In addition, the secular resonant angle $\Delta\omega$ = $\varpi_{\rm b}$ - $\varpi_{\rm c}$ = $\theta_1$ - $\theta_2$
is an important indicator for a libration in a secular resonance.

\begin{figure*}[btp]
%\begin{center}$
%\begin{array}{c} 
\sidecaption
%\centering
\includegraphics[width=10.1cm]{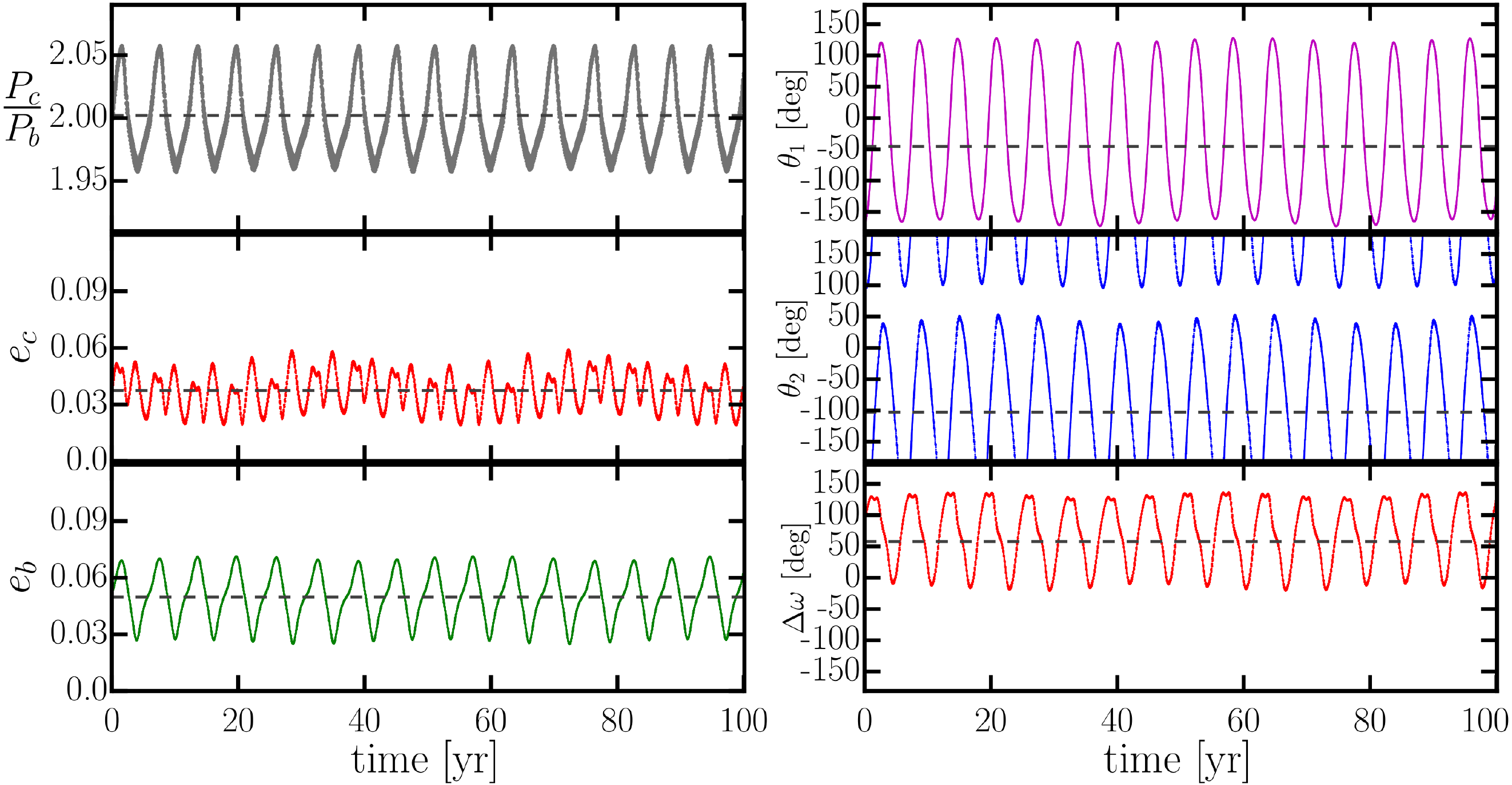}
%\end{array} $
%\end{center}
\caption{
One-hundred year  time interval of the orbital evolution of HD~27894 b and c, with initial eccentricities $e_{\rm b}$ = 0.052 and $e_{\rm c}$ = 0.035 (within the formal uncertainties of the 3PDM).
In this configuration the planetary eccentricity $e_{\rm b}$ oscillates between 0.02 and 0.07, and $e_{\rm c}$ between 0.02 and 0.06,
with mean values of 0.050 and 0.037, respectively.
The resonance angles
$\theta_1 \approx -50^\circ$, $\theta_2 \approx -100^\circ$, and $\Delta\omega \approx 50^\circ$ 
are oscillating with large amplitudes in a clear asymmetric 2:1 MMR.
}
\label{evol} 
\end{figure*}

We examined the orbital evolution of $\theta_1,\theta_2$, and $\Delta\omega$ of the 3PDM 
and we find that all three angles circulate from 0$^\circ$ to 360$^\circ$, indicating a 
regular secular motion with a 2:1 period ratio, but no explicitly resonant motion.
According to \citet{Lee2004} a stable 2:1 MMR configuration with both $\theta_1$ and $\theta_2$ librating 
can be achieved by smooth migration capture.
For cases where a pair of planets with mass ratio of 2.75 $\lesssim$ $m_{\rm b}$/$m_{\rm c}$ $\lesssim$ 5 
migrates with sufficiently small eccentricities, \citet{Lee2004} suggests that
$\theta_1$ and $\theta_2$ could in principle be trapped in an anti-aligned libration leading to $\Delta\omega$ = 180$^\circ$,
or in the case of larger eccentricities at any angle exhibiting an asymmetric 
2:1 MMR  \citep[see also][]{Beauge2003, FerrazMello2003}.

Therefore, with the aim of understanding how the dynamical properties of planets b and c depend on their eccentricities
and orbital alignment, we studied a large number of three-planet configurations similar to the 3PDM.
We systematically varied $e_{\rm b}$, $e_{\rm c}$, $\omega_{\rm b}$, $\omega_{\rm c}$, $M_{\rm b}$,  and $M_{\rm c}$ 
within their 1$\sigma$ bootstrap confidence levels, while keeping the remaining parameters fixed at the 3PDM best-fit values. 
We integrated these configurations for 10 k\,yrs, and we evaluated whether the resonance angles
$\theta_1,\theta_2$, and $\Delta\omega$ were librating.
We found that when $e_{\rm b}$ and $e_{\rm c}$ are very small,
the two planets exhibit an anti-aligned resonance 
with $\theta_1$ $\approx$ 0$^\circ$, $\theta_2$ $\approx$ 180$^\circ$, and $\Delta\omega$ $\approx$ 180$^\circ$, 
which is expected if planets b and c have been trapped in a 2:1 MMR during migration with near zero eccentricities.
However, when $e_{\rm b}$ and $e_{\rm c}$ are close to their 1$\sigma$ upper limits, we found
configurations with $\theta_1$ $\approx -50^\circ$, $\theta_2 \approx -100^\circ$, and $\Delta\omega$ $\approx 50^\circ$,
which is an asymmetric 2:1 MMR, 
as predicted by \citet{Beauge2003}, \citet{FerrazMello2003}, and \citet{Lee2004}. 
Perhaps this is the more likely scenario,
since the two giant planets can have somewhat larger eccentricities during migration. 
A 100~yr time interval illustrating the evolution of such an asymmetric 2:1 MMR is shown in Fig.~\ref{evol}.
This configuration has $e_{\rm b}$ = 0.052 and $e_{\rm c}$ = 0.035 initially, which is within the formal uncertainties of the 3PDM
and thus represents a possible configuration of HD~27894's inner planet pair.

\vspace{-0.15cm}

\section{Discussion}
\label{sec4}

The co-existence of a massive, distant, and moderately eccentric Jovian planet 
and a possibly resonant inner pair of massive planets makes the HD~27984 system truly unique.  
The orbital configuration may suggest that planet formation and evolution
depends strongly on the disk properties and migration rate of the planets during the protoplanetary disk stage.
The inner pair of planets has probably been trapped in a 2:1 
MMR by smooth migration capture, while the outer massive Jovian planet migrated 
to 5.4 AU before the circumstellar disk evaporated, thus halting the migration.

There are some strong indications that the inner two planets are in a 2:1 MMR, although so far
we have been unable to definitely 
%\LEt{ to incontrovertibly confirm, to confirm with certainty ?   } 
confirm the resonance based on the best-fit dynamics.
For the formal best fit, the mean period ratio of the inner planets is very close to 2:1,
but the resonance angles $\theta_1$ and $\theta_2$ circulate from 0$^\circ$ to 360$^\circ$.
Our first attempt to find a stable librating solution within the 
1$\sigma$ confidence region of our 3PDM, however, was successful.
We identified resonant configurations where we either see anti-aligned resonances with 
$\Delta\omega \approx 180^\circ$, or
an asymmetric configuration where $\Delta\omega$ $\approx 50^\circ$ and $\theta_1$ $\approx -50^\circ$,
$\theta_2$ $\approx -100^\circ$.
To our knowledge, HD~27894 is the first system to show signs of an asymmetric 2:1 MMR
in agreement with the migration theory for a mass ratio of 4 $\lesssim$ $m_{\rm b}$/$m_{\rm c}$.

Nevertheless, more work and more data are needed to understand and resolve the possible 
resonance between planets b and c, as well as its origin. 
The current data allowed us to conclusively confirm the presence of three planets,
but more data are needed to confirm the resonant nature 
of the inner planets and to further constrain the orbital parameters, especially of HD~27894~d. 
 The outermost planet is currently sparsely covered by the RV data and 
thus we cannot exclude the possibility that its true $P_{\rm d}$, $e_{\rm d}$, and $m_{\rm d} \sin i$ 
are not within our 3PDM best-fit uncertainties.
We plan to continue our HARPS monitoring of HD~27894, which will allow us to extend our 
orbital RV analysis using a multidimensional parameter grid-search 
or Bayesian framework for parameter estimation, and to carry out a more conclusive long-term dynamical analysis.

The planetary system around HD~27894 is important for probing planetary formation and evolution scenarios
and illustrates the importance of further follow-up of RV planet hosts.

\begin{acknowledgements}
We thank Megan Bedell and Oscar Barragan for observations of this target in the framework of observing time exchange.
We also thank Gaspare Lo Curto for very useful discussions during the course of our survey.
M.Z. has received financial support from the DFG under RE 1664/12-1.
Z.O.V.\ acknowledges a German Academic Exchange Service (DAAD) stipend (91613113).
 M.H.L. was supported in part by the Hong Kong RGC grant HKU 7024/13P.
This work used the python package astroML \citep{VanderPlas2012} for the calculation of the GLS periodogram.
We thank the anonymous referee for the constructive comments that helped to improve this paper.

\end{acknowledgements}

\vspace*{-0.4cm}

\bibliographystyle{aa}

\bibliography{hd27894_bib} 

\begin{appendix} %First online appendix]

 \setcounter{table}{0}
\renewcommand{\thetable}{A\arabic{table}}

 \setcounter{figure}{0}
\renewcommand{\thefigure}{A\arabic{figure}}

\begin{table}
\caption{HARPS radial velocity measurements for HD~27894} 
\label{table:HD27894} 

\centering  

\begin{tabular}{c r c } 

\hline\hline    
\noalign{\vskip 0.5mm}

epoch [JD] & RV [m\,s$^{-1}$] & $\sigma_{RV}$ [m\,s$^{-1}$]  \\

\hline
\noalign{\vskip 0.9mm}
2452941.827   &   36.31   &    2.11  \\ 
2453273.878   &   -0.79   &    1.22  \\ 
2453274.882   &   -23.38   &    1.47  \\ 
2453289.859   &   50.42   &    1.01  \\ 
2453294.827   &   -41.49   &    1.27  \\ 
2453297.861   &   -28.44   &    1.51  \\ 
2453340.762   &   67.73   &    1.19  \\ 
2453341.705   &   62.97   &    1.04  \\ 
2453342.716   &   49.98   &    1.32  \\ 
2453344.742   &   17.19   &    1.08  \\ 
2453345.656   &   -1.13   &    1.65  \\ 
2453366.674   &   -47.77   &    0.69  \\ 
2453369.685   &   -35.97   &    0.97  \\ 
2453371.684   &   -0.24   &    1.71  \\ 
2453372.643   &   18.71   &    0.56  \\ 
2453374.682   &   49.47   &    1.42  \\ 
2453375.682   &   57.14   &    1.27  \\ 
2453376.626   &   64.64   &    1.23  \\ 
2453377.617   &   62.71   &    1.39  \\ 
2453378.654   &   56.14   &    1.33  \\ 
2453400.626   &   -22.37   &    1.10  \\ 
2453403.583   &   -53.41   &    1.03  \\ 
2453404.609   &   -49.29   &    0.95  \\ 
2453410.578   &   38.66   &    1.58  \\ 
2453974.903   &   -43.94   &    1.65  \\ 
2453974.906   &   -41.39   &    1.69  \\ 
2453974.909   &   -42.84   &    1.74  \\ 
2453980.888   &   -123.90   &    1.32  \\ 
2453980.892   &   -122.24   &    1.16  \\ 
2453980.895   &   -121.65   &    1.50  \\ 
2453981.906   &   -112.93   &    1.11  \\ 
2453981.909   &   -113.22   &    1.39  \\ 
2453981.912   &   -111.33   &    1.18  \\ 
2453982.891   &   -99.47   &    1.21  \\ 
2453982.895   &   -99.29   &    1.05  \\ 
2453982.898   &   -100.80   &    1.16  \\ 
2454049.769   &   -103.54   &    0.74  \\ 
2454051.757   &   -123.05   &    0.95  \\ 
2454053.769   &   -117.39   &    0.68  \\ 
2454055.753   &   -87.32   &    0.68  \\ 
2454055.763   &   -87.00   &    0.71  \\ 
2456624.601   &   -155.17   &    0.70  \\ 
2457609.896   &   -20.42   &    1.22  \\ 
2457609.902   &   -21.12   &    1.11  \\ 
2457609.908   &   -20.26   &    1.19  \\ 
2457610.870   &   -11.04   &    1.22  \\ 
2457610.881   &   -11.02   &    1.23  \\ 
2457610.893   &   -11.08   &    1.17  \\ 
2457682.836   &   -6.60   &    1.07  \\ 
2457682.842   &   -4.47   &    0.96  \\ 
2457682.848   &   -6.57   &    1.09  \\ 
2457686.836   &   -33.88   &    1.05  \\ 
2457686.842   &   -35.17   &    1.05  \\ 
2457686.848   &   -35.54   &    1.35  \\ 
2457689.869   &   -92.08   &    1.18  \\ 
  
\hline       

\end{tabular}

\tablefoot{
Velocities taken after JD = 2457609.896 were 
adjusted by $-$13.4 m\,s$^{-1}$ to correct for the 
induced RV offset due to the HARPS fiber upgrade  \citep[May 2015; see][and text for details]{LoCurto2015}. % See text for details.
}

\end{table}

\end{appendix}

\end{document}